\documentclass[aps,amsmath,amssymb,prd,superscriptaddress,nofootinbib]{revtex4}
\usepackage{amsmath}
\usepackage{bm}
\usepackage[dvips]{graphicx}
\newcommand{\be}{\begin{equation}}

\newcommand{\ee}{\end{equation}}
\newcommand{\bea}{\begin{eqnarray}}
\newcommand{\eea}{\end{eqnarray}}

\begin{document}

\title{Three flavor Nambu-Jona Lasinio model
with Polyakov loop and competition with nuclear matter}

\author{M. Ciminale}\email{marco.ciminale@ba.infn.it}
\affiliation{Universit\`a di Bari, I-70126 Bari, Italia}
\affiliation{I.N.F.N., Sezione di Bari, I-70126 Bari, Italia}
\author{R.Gatto}\email{raoul.gatto@physics.unige.ch}
\affiliation{D\'epartement de Physique Th\'eorique, Universit\'e
de Gen\`eve, CH-1211 Gen\`eve 4, Suisse}
\author{N. D. Ippolito}\email{nicola.ippolito@ba.infn.it}
\affiliation{Universit\`a di Bari, I-70126 Bari, Italia}
\affiliation{I.N.F.N., Sezione di Bari, I-70126 Bari, Italia}
\author{G. Nardulli}\email{giuseppe.nardulli@ba.infn.it}
\affiliation{Universit\`a di Bari, I-70126 Bari, Italia}
\affiliation{I.N.F.N., Sezione di Bari, I-70126 Bari, Italia}
\author{M. Ruggieri}\email{marco.ruggieri@ba.infn.it}
\affiliation{Universit\`a di Bari, I-70126 Bari, Italia}
\affiliation{I.N.F.N., Sezione di Bari, I-70126 Bari, Italia}

\begin{abstract}
We study the phase diagram of the three flavor Polyakov-Nambu-Jona
Lasinio (PNJL) model and in particular the interplay between chiral
symmetry restoration and deconfinement crossover. We compute chiral
condensates, quark densities and the Polyakov loop at several values
of temperature and chemical potential. Moreover we investigate on
the role of the Polyakov loop dynamics in the transition from
nuclear matter to quark matter.
\end{abstract}
 \preprint{BA-TH 5YZ/07}

\maketitle

\section{Introduction}
A long-standing problem with Quantum Chromodynamics (QCD) is the
difficulty of an analytical study in the nonperturbative regime. To
deal with this issue a few approximate approaches have been
developed, based on the idea of effective field theories.
 Among them, the Nambu-Jona Lasinio model~\cite{Nambu:1961tp}
 (NJL in the following) has become a
popular tool to describe some of the salient aspects of low energy
QCD, in particular chiral symmetry breaking and its restoration at
high density and/or temperature (see Refs.~\cite{revNJL} for
reviews).

The NJL model neglects the gluon dynamics, the interactions among
quarks being described by contact terms. Due to the absence of
gluons, one of the main properties of low-energy QCD at small
temperature and baryonic density, namely confinement, is missing in
this effective description. This defect not-withstanding, the NJL
model enjoys a significant popularity since it allows analytical
treatments in different contexts.  In order to cure its deficiencies
some extensions have been also proposed. They either try to describe
quarks and hadrons in unified approaches (see
Ref.~\cite{Lawley:2006ps} and references therein), or add a bag
constant to the equation of state of the NJL quark
matter~\cite{Wang:2006fh}. In the latter procedure the deconfinement
transition is obtained by computing the pressure of nuclear matter
at low density and temperature by some effective model and comparing
it to the pressures of the quark matter.  When the nuclear matter pressure
becomes smaller than the quark pressure, deconfinement is
energetically favored. We stress that the introduction of a bag
constant is needed to reproduce the phase transition from the
nuclear to the quark phase in this approach.

Another  extension of the NJL model has been suggested in
Ref.~\cite{Fukushima:2003fw}. In this paper part of the gluon
dynamics is described by a background temporal gluon field coupled
to quarks  by the QCD covariant derivative. The background field
adds a potential term ${\cal U}(\Phi)$ to the lagrangian. Its value
is determined for any temperature $T$ and quark chemical potential
$\mu$ by the minimization of the effective potential and depends on
the traced Polyakov loop $\Phi$~\cite{Polyakovetal}, related to the
background gauge field. The resulting model is known as
Polyakov-Nambu-Jona Lasinio (PNJL) model.

As is well known the Polyakov loop in a pure gauge theory is an
order parameter of the deconfinement transition~\cite{Polyakovetal}.
This peculiarity is related to the existence of a discrete symmetry
$Z_3$ of the pure gauge action, which is spontaneously broken when
deconfinement sets in. The Polyakov loop vanishes in the disordered
low temperature phase and is different from zero in the high $T$
phase. When dynamical quarks are added, $Z_3$ symmetry is explicitly
broken and $\Phi$ can no longer be considered as an order parameter.
However, as shown by lattice simulations, its behavior as a function
of $T$ (increasing from zero to non vanishing values when $T$
increases) can  still describe the deconfinement crossover. One
might assume that this happens also when the chemical potential
$\mu$ is varied, which should show a link between deconfinement and
chiral restoration~\cite{linking,Fukugitaetal}. Therefore in the
PNJL model, one introduces a Polyakov loop dynamics in the NJL model
trying to reproduce both chiral symmetry breaking and quark
confinement.

The PNJL model with two flavors has been extensively
studied~\cite{Ratti:2005jh,Roessner:2006xn,Ghosh:2007wy,
Kashiwa:2007hw,Schaefer:2007pw,Ratti:2007jf,Sasaki:2006ww,Megias:2006bn}.
In Ref.~\cite{Ciminale:2007ei} an extension to the three flavor
model has been considered, by studying the effects of the Polyakov
loop on the thermodynamics of the Color-Flavor-Locked phase of
high density QCD~\cite{Alford:1998mk} near its second order
transition to the normal phase. The aim of this paper is to extend
this analysis to moderate densities. In particular we will study
the phase diagram of the three flavor PNJL model and the
transition from nuclear to deconfined quark matter. We shall
consider massive quarks because, differently from previous
work~\cite{Ciminale:2007ei}, mass effects cannot be neglected at
moderate values of the quark chemical potential. Therefore we
introduce bare quark masses and also compute selfconsistently
their in-medium values in the mean field approximation. On the
other hand, since we are mainly interested in the interplay of
chiral symmetry restoration  and  deconfinement crossover we
neglect the possibility of color superconductivity, which should
be produced at higher
densities~\cite{CSC,Ruster:2005jc,Blaschke:2005uj,Ippolito:2007uz}. While chiral
symmetry restoration is adequately described by the PNJL
model, to establish the deconfinement transition we have to
consider also nuclear matter, described by an effective field
theory, so that a comparison of free energies can be made.

The plan of the paper is as follows. In Section II we summarize the
main features of the three flavor PNJL model. Section III is devoted
to the study of the phase diagram of the model. In Section IV we
compare PNJL quark matter with nuclear matter. Finally, in Section V
we draw our conclusions.

\section{Thermodynamics of the three flavor PNJL
model}\label{Sec:Thermo} We consider the lagrangian\begin{equation}
{\cal L} = \sum_f\bar\psi_f\left(\,iD_\mu \gamma^\mu - m_f +
\mu\gamma_0\right)\psi_f ~+~{\cal L}_4~+~{\cal
L}_6~,\label{eq:lagr1}
\end{equation}
where the sum is over the three flavors $f$ $(=1,2,3$ for $u,d,s$).
In the above equation we have introduced the coupling of the quarks
to a background gauge field $A_\mu = g\delta_{\mu0}A_{a\mu}T_a$ via
the covariant derivative $D_\mu = \partial_\mu-i A_\mu$; $m_f$ is
the current mass (we assume $m_u = m_d$). The quark chemical
potential is denoted by $\mu$. The NJL four-fermion and six-fermion
interaction Lagrangians are as follows \cite{revNJL}:\bea {\cal L}_4
&=& G\sum_{a=0}^8\left[\left(\bar\psi \lambda_a \psi\right)^2 +
\left(i\bar\psi \gamma_5\lambda_a \psi\right)^2
\right]\label{eq:full4}~,\\ {\cal L}_6 &=& -K\left[{\rm
det}\bar\psi_f(1+ \gamma_5)\psi_{f'} \,+\,{\rm det}\bar\psi_f(1-
\gamma_5)\psi_{f'} \right]\ , \label{eq:full6} \eea where
$\lambda_a$ are the Gell-Mann matrices in flavor space ($\lambda_0 =
\sqrt{2/3}~{\bm 1}_f$) and the determinant is in flavor space as
well. Working in the mean field approximation the self-consistent
equations for the constituent quark masses are
\begin{equation}
M_f = m_f - 4 G \sigma_f~+2\,\,K\,\sigma_{f+1}\,\sigma_{f+2}~;
\label{eq:masseMU0}
\end{equation}
here $\sigma_f = \langle\bar{f}f\rangle$ denotes the chiral
condensate of the flavor $f$, and we define
$\sigma_4=\sigma_u,\,\sigma_5=\sigma_d$. We also introduce the
quark mass matrix $M=\text{diag}[M_u,M_d,M_s]$. The gap equation
at $T=0$ and $\mu=0$ is \be
\sigma_f=-\frac{3M_f}{\pi^2}\int_0^\Lambda\frac{p^2}{\sqrt{p^2+M_f^2}}
\,dp~, \ee  which depends on the ultraviolet cutoff $\Lambda$. The
parameters are chosen as in Ref.~\cite{Ruster:2005jc}, i.e. \be
m_{u,d}\ = \ 5.5~\text{MeV}~,~m_{s} \ = \ 140.7~\text{MeV}~,\
G\Lambda^2 = 1.835~,~ K\Lambda^5= 12.36~, ~\Lambda =
602.3~\text{MeV}~. \ee By these parameters one gets $m_\pi \simeq
135$ MeV, $m_K \simeq 498$ MeV, $m_{\eta^\prime} \simeq 958$ MeV,
$m_\eta \simeq 515$ MeV and $f_\pi \simeq 92$ MeV.

Once the lagrangian is specified, the thermodynamic potential at
temperature $T$ is obtained after integration over the fermion
fields in the partition function:\be \Omega = {\cal U}[T,\Phi,\bar\Phi]
+2G\sum_{f=u,d,s}\sigma_f^2 - 4 K \sigma_u \sigma_d \sigma_s - T
\sum_n
\int\frac{d^3p}{(2\pi)^3}\text{Tr}\log\frac{S^{-1}(i\omega_n,{\bm
p})}{T}~.\label{eq:due} \ee Here $\omega_n = \pi T (2n+1)$ are
Matsubara frequencies. The inverse quark propagator is given in
momentum space by
\begin{equation}
S^{-1} = \gamma_0 \left(p^0 +\mu -
iA_4\right)-{\bm\gamma}\cdot{\bm p} - M~,
\end{equation}
where $A_4 = i A_0$. A most significant difference between the NJL
and the PNJL model is the presence in the thermodynamic potential
of the gluon contribution ${\cal U}(T,\Phi,\bar\Phi)$ describing
the dynamics of the traced  Polyakov loop
\begin{equation}
\Phi = \frac{1}{3}\text{Tr}_c {\cal P} \exp\left[i\int_0^\beta
d\tau A_4({\bm x},\tau)\right]~,~~~~~\bar\Phi =
\frac{1}{3}\text{Tr}_c {\cal P} \exp\left[-i\int_0^\beta d\tau
A_4^\star({\bm x},\tau)\right]~,
\end{equation} in absence of dynamical quarks.
We will consider in the following two different choices of the
potential. First we assume a polynomial expansion in $\Phi$ and
$\bar\Phi$:
\begin{equation}{\rm Polynomial~model:}~~~~~~
\frac{{\cal U}[T,\Phi,\bar\Phi]}{T^4}=
-\frac{b_2(T)}2\Bar{\Phi}\Phi\, - \frac{b_3}{6}\left(\Phi^3 +
\bar\Phi^{3}\right) + \frac{b_4}{4}(\Bar{\Phi}
\Phi)^2~,\label{eq:Pot1}~~~~~~~~~~~~
\end{equation}
where \be
b_2(T)=a_0+a_1\left(\frac{T_0}{T}\right)+a_2\left(\frac{T_0}{T}\right)^2
+a_3\left(\frac{T_0}{T}\right)^3~.\ee This corresponds to the
choice $\theta=0$ for the parameter $\theta$ introduced in
\cite{Schaefer:2007pw}.  Numerical values of the coefficients
are~\cite{Ratti:2005jh} \be
a_0=6.75,~~a_1=-1.95,~~a_2=2.625,~~a_3=-7.44~,~~b_3=0.75~,~~b_4=7.5~.
\ee The remaining parameter $T_0$ is the deconfinement temperature
of static and infinitely heavy quarks~\cite{Polyakovetal}. This
parameter depends in principle on the number of flavors and on the
quark chemical potential. We shall investigate two cases. First,
we assume a constant value $T_0=270$ MeV, which corresponds to the
deconfinement temperature of static and infinitely heavy quarks as
computed in lattice QCD (see for example~\cite{Roessner:2006xn}).
The second case we study is the $\mu$-dependent $T_0$ suggested
in~\cite{Schaefer:2007pw},
\begin{equation}
T_0 = T_\tau e^{-1/(\alpha_0 b(\mu))}~, \label{eq:TTT}
\end{equation}
with $T_\tau = 1770$ MeV, $\alpha_0 = 0.304$ and
\begin{equation}
b(\mu) = \frac{1}{6\pi}(11N_c - 2N_f) - \frac{16}{\pi}N_f
\frac{\mu^2}{T_\tau^2}
\end{equation}
(this dependence is  motivated by the use of hard dense loop and
hard thermal loop results for the effective charge
\cite{Schaefer:2007pw,lebellac}). The prescription in
(\ref{eq:TTT}) gives $T_0 = 178$ MeV at $\mu=0$.

The second model we consider uses instead of (\ref{eq:Pot1}) the
following logarithmic form \be{\rm
Logarithmic~model:}~~~~~~\frac{{\cal U}(T,\Phi,\bar\Phi)}{T^4}=
-\frac{\tilde
b_2(T)}2\Bar{\Phi}\Phi\,+\,b(T)\ln[1-6\Bar{\Phi}\Phi+4(
\Phi^3+\bar\Phi^{3})-(\Bar{\Phi}\Phi)^2]\ee with\be \tilde
b_2(T)=\tilde a_0+\tilde a_1\left(\frac{T_0}{T}\right)+\tilde
a_2\left(\frac{T_0}{T}\right)^2\,,~~~b(T)=\tilde b_3
\left(\frac{T_0}T\right)^3\,.\ee Numerical values of the
coefficients are as follows~\cite{Roessner:2006xn}\be \tilde a_0=3.51,\,\tilde
a_1=-2.47,\,\tilde a_2=15.2,\,\tilde b_3=-1.75\,.\ee In this
second case we only consider the case of a fixed value of $T_0$:
$T_0$=183 MeV, corresponding to the $N_f=3$ value of the
deconfinement temperature estimated in
Ref.~\cite{Schaefer:2007pw}.

Evaluation of the trace gives in both cases for the thermodynamic
potential the expression
\begin{eqnarray}
\Omega &=& {\cal U}[T,\Phi,\bar\Phi] +2G\sum_{f=u,d,s}\sigma_f^2 - 4 K
\sigma_u \sigma_d \sigma_s \,-\,
6\sum_{f=u,d,s}\int\frac{d^3p}{(2\pi)^3}E_f~\theta(\Lambda-|{\bm
p}|) \nonumber \\
&&-2T\sum_{f=u,d,s}\int\frac{d^3p}{(2\pi)^3}\log\left[1 + 3\Phi
e^{-\beta(E_f - \mu)} + 3\Bar{\Phi} e^{-2\beta(E_f - \mu)} +
e^{-3\beta(E_f - \mu)} \right]
\nonumber \\
&&-2T\sum_{f=u,d,s}\int\frac{d^3p}{(2\pi)^3}\log\left[1 +
3\Bar{\Phi} e^{-\beta(E_f + \mu)} + 3\Phi e^{-2\beta(E_f + \mu)} +
e^{-3\beta(E_f + \mu)} \right]~,\label{eq:OmegamupNJL}
\end{eqnarray}
with $E_f = \sqrt{{\bm p}^2 + M_f^2}$ and $M_f$ given by
Eq.~\eqref{eq:masseMU0}. By searching the global minimum of $\Omega$
one can get quark condensates $\sigma_f$ and Polyakov loop for any
values of  the parameters $\mu$ and $T$.

\section{Numerical results\label{Sec:Numerical}}

\begin{figure}[h]
\begin{center}
\includegraphics[width=11cm]{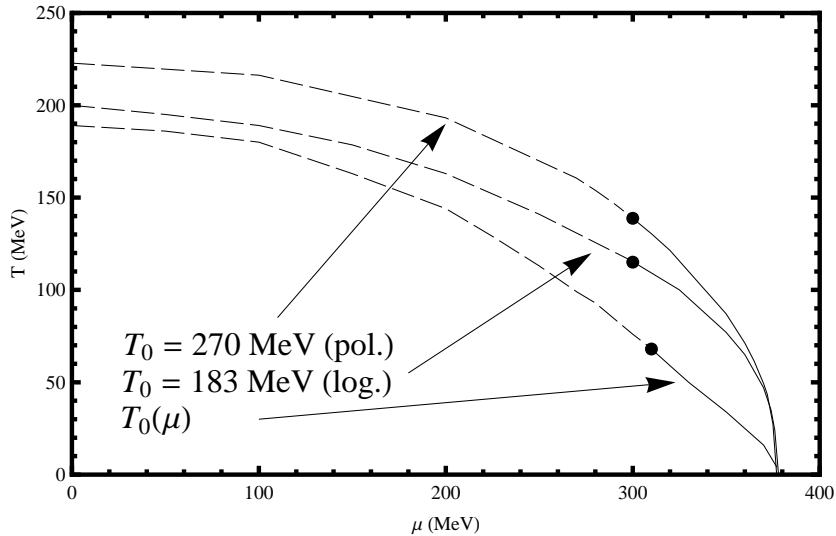}
\end{center}
\caption{\label{Fig:diagramma} Phase diagram of the PNJL model.
The three lines correspond to the following models: $T_0$ fixed at
the constant value 270 MeV and polynomial form for the Polyakov
potential ${\cal U}(\Phi)$ (highest); $T_0=183$ MeV and
logarithmic form of the potential (middle); $T_0=T_0(\mu)$ and
polynomial form (lowest). A solid line denotes a first order
chiral transition, a dashed line a cross-over for the condensate
$<\bar uu>$ and a bold point a critical endpoint (CEP).}
\end{figure}

In Fig.~\ref{Fig:diagramma} we show the phase diagrams of three
cases mentioned above (two cases for the polynomial model, one for
the logarithmic model). They are obtained as follows. For any
value of the quark chemical potential, the critical temperature is
identified with the temperature at which the up quark chiral
condensate $\sigma_u$ has a jump (first order transition), or the
derivative of $\sigma_u$ with respect to the temperature is
maximum (second order transition).
The phase diagrams of the polynomial model correspond to the
highest and lowest lines, while the intermediate line is obtained
by the logarithmic potential. The two lines of the polynomial
model are obtained with the fixed value $T_0=270$ MeV (upper line)
and with $T_0(\mu)$ of Eq. (\ref{eq:TTT}) (lower) respectively.
Qualitatively the three curves are similar, which shows the
physical results are insensitive to the details of the model.  The
critical endpoints (CEPs) separating the cross-over from the
first-order lines are located, for the three cases, at $(\mu_E,
T_E) \approx (300,140)$ MeV (highest curve), $(\mu_E, T_E) \approx
(300,115)$ MeV (middle curve) $(\mu_E, T_E) \approx (310,68)$ MeV
(lowest curve).  At $\mu=0$ we find for the three cases the
cross-over temperatures $T_\chi(\mu=0)\approx 223$ MeV (highest
curve), $T_\chi(\mu=0)\approx 200$ MeV (middle curve),
$T_\chi(\mu=0)\approx 189$ MeV (lowest curve).

The numerical values of the cross-over temperatures at $\mu=0$ are
in the same range of the results found in QCD-like
theories~\cite{Kiriyama:2000yp,Kiriyama:2001ah,Hashimoto:2005pd}
or in lattice QCD~\cite{LATTICE}. On the other hand, the CEPs are
located at values of $\mu$ much higher than the respective values
found in QCD/like
theories~\cite{Kiriyama:2000yp,Kiriyama:2001ah,Hashimoto:2005pd},
lattice QCD~\cite{LATTICE} and empirical analysis of the ratio of
shear viscosity to entropy density~\cite{Lacey:2007na}, which
suggest there is a CEP at $T\approx165$ MeV and $\mu\approx50$
MeV. Clearly the details of the models matter; in relation to this
problem, in Ref.~\cite{Kashiwa:2007hw} an extension of the PNJL
model (in the case of two flavors) has been proposed using also a
${\cal O}(\bar\psi \psi)^4$ interaction term. It was found that
the effect of the new kind of interaction is to locate the
critical end point at higher temperature and lower chemical
potential than the NJL/PNJL results. One might expect similar
effects in the three flavor models considered in the present
paper.
\begin{figure}[h]
\begin{center}
\includegraphics[width=7cm]{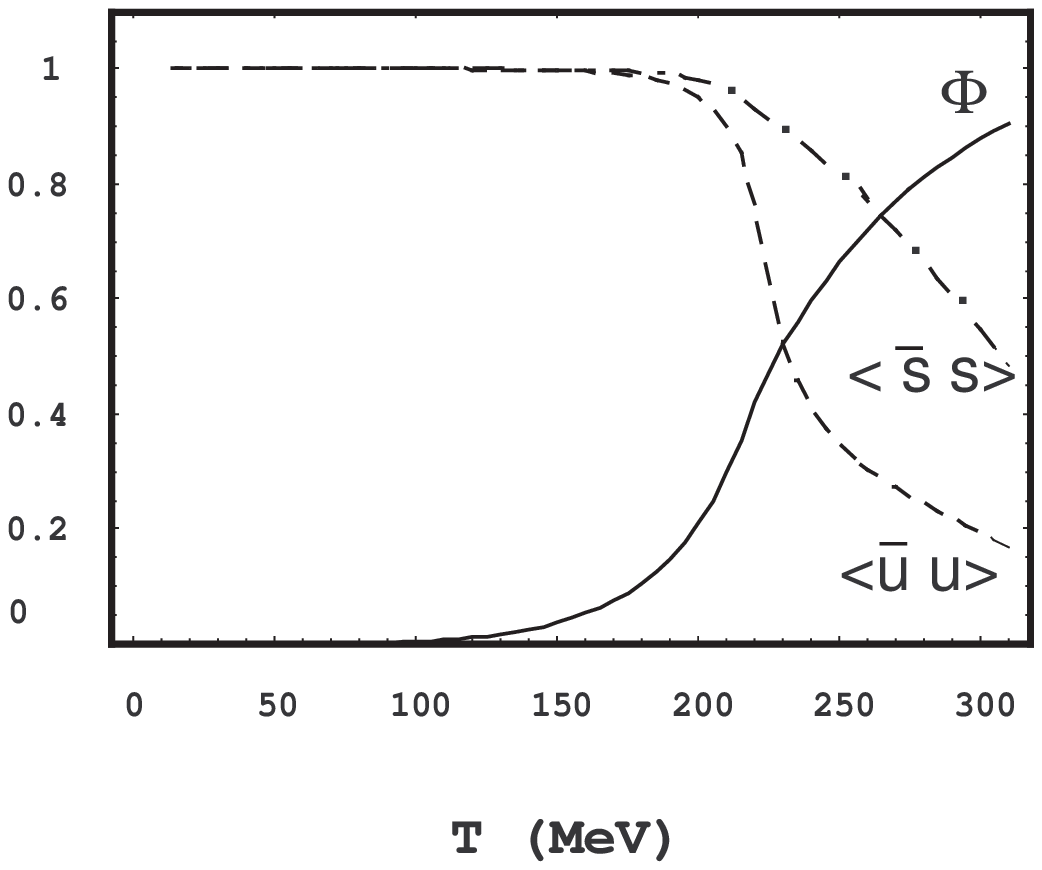}~~~~~~~~
\includegraphics[width=7.1cm]{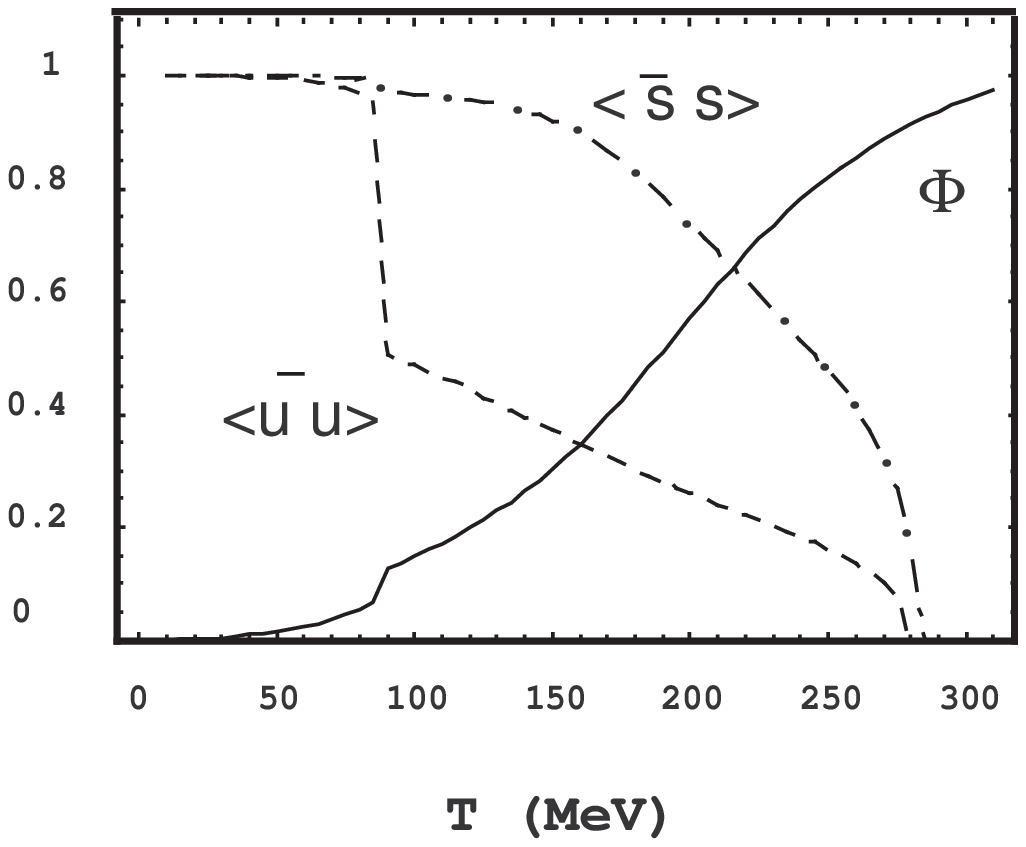}\\
~~\includegraphics[width=7.3cm]{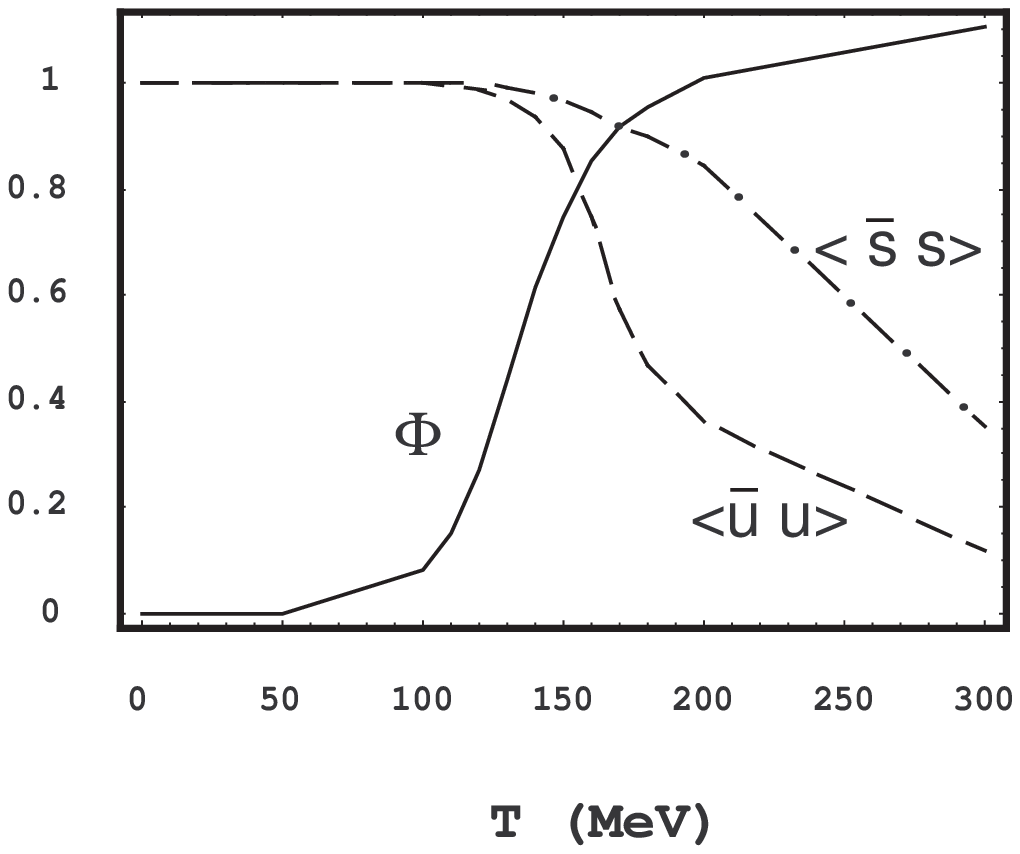}
~~\includegraphics[width=7.6cm]{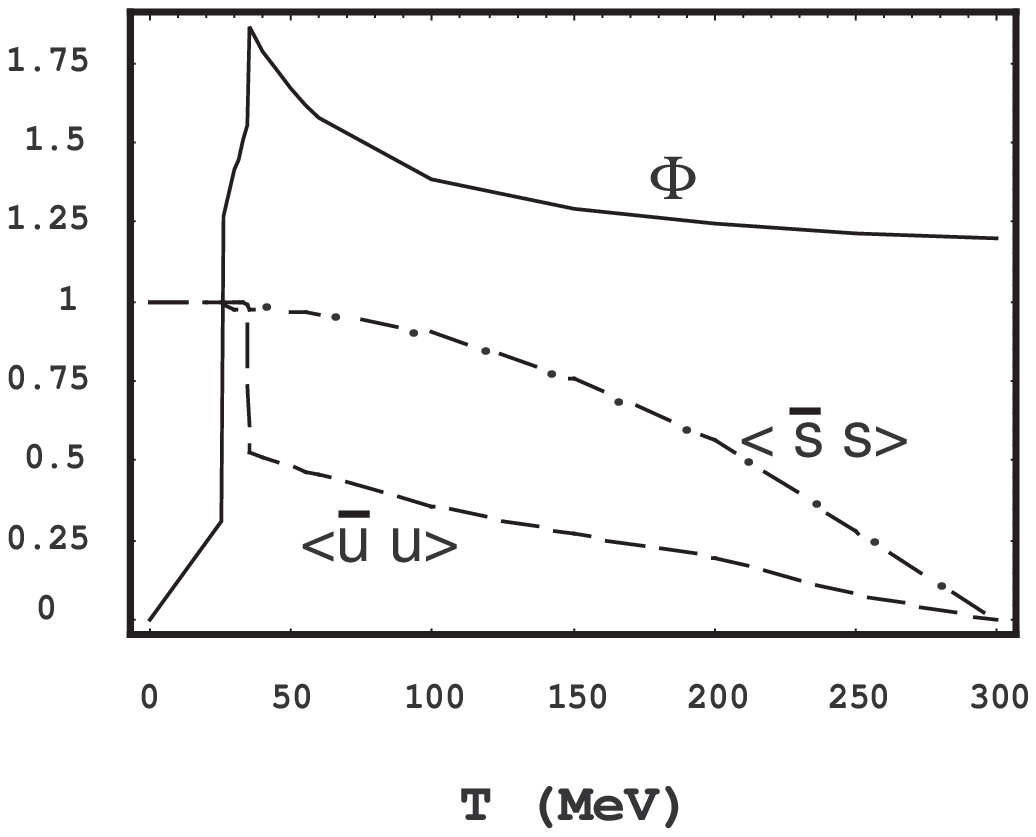} \caption{\label{Fig:Fig2}
Upper panels: Polyakov loop $\Phi$ (solid line) and chiral
condensates for $u$ quark (dashed line) and $s$ quark (dot-dashed
line) versus temperature in MeV, at $\mu=150$ MeV (left) and
$\mu=350$ MeV (right), for the PNJL model with fixed $T_0 = 270$
MeV. For each value of $\mu$ the condensates are normalized to their
values at $T=0$. Lower panels: Same quantities for the model with
$T_0 = T_0(\mu)$.}\end{center}
\end{figure}

The phase diagrams in Fig.~\ref{Fig:diagramma} are based on the
results of  Fig.~\ref{Fig:Fig2} that shows the $T-$ dependence (at
fixed $\mu$) of the chiral condensates. The $u$ chiral condensates
are represented by dashed lines, those of the strange quark by
dot-dashed lines (solid lines represent the Polyakov loop $\Phi$).
We have reported results only for the polynomial model for the two
cases of fixed $T_0=270$ MeV (upper panels) and $T_0(\mu)$ (lower
panels), since the results for the case of the logarithmic
potential are qualitatively similar to those of the case of the
polynomial potential with $T_0 = 270$ MeV. On the right panel of
Fig.~\ref{Fig:Fig2} we have results at $\mu=350$ MeV, on the left
those at $\mu=150$ MeV.

From the diagrams of Fig.~\ref{Fig:Fig2} we can identify another
relevant transition temperature: $T_\Phi$, i.e. the temperature
corresponding to the inflection point of $\Phi$ (defined as the
temperature where $ d\Phi/dt$ has a maximum). As discussed in the
introduction $T_\Phi$ cannot be immediately interpreted as the
deconfinement temperature because $Z_3$ is not a symmetry of QCD
with dynamical quarks. In fact in the next section we shall adopt
a different method to study the deconfinement transition, based on
the comparison of the pressures of the two competing states of
matter (quark and nuclear). Nevertheless  $T_\Phi$ should
approximately coincide with the deconfinement transition, if the
guess on the relevance of the rise of Polyakov loop for the
deconfinement of dynamical quarks is correct.  One can note a
different behavior between the two cases $T_0=270$ MeV (fixed) and
$T_0(\mu)$. In the former case $T_\Phi$ numerically coincides
either with the chiral crossover (at low $\mu$) or with the chiral
first order phase transition (high $\mu$). In the latter case this
equality is lost, the shift between $T_\phi$ and $T_\chi$ being of
the order of $30$ MeV for $\mu$ in the interval $[0,320]$ MeV, and
shrinking as $\mu$ is increased above $320$ MeV. Another
peculiarity of the $T_0(\mu)$ case is that at high temperatures
$\Phi\,>\,1$. This is not acceptable since $\Phi$ is defined as
the normalized trace of a $SU(3)$ matrix and should therefore lie
in the interval $0\,\le\,|\Phi|\,\le\,1$. The fact that $|\Phi|$
can be larger than one in the PNJL model with a polynomial
potential is well known in the two flavor
case~\cite{Ratti:2005jh}, and to solve this problem several
improvements have been suggested, e.g the use of the logarithmic
potential~\cite{Fukushima:2003fw,Roessner:2006xn} or the
interpretation of the Polyakov loop potential as a random matrix
model~\cite{Ghosh:2007wy,Dumitru:2003hp,Dumitru:2005ng}. In our
numerical calculations within the logarithmic potential model we
have checked that $|\Phi|$ is always less than one.

\begin{figure}[h]
\begin{center}
\includegraphics[width=14cm]{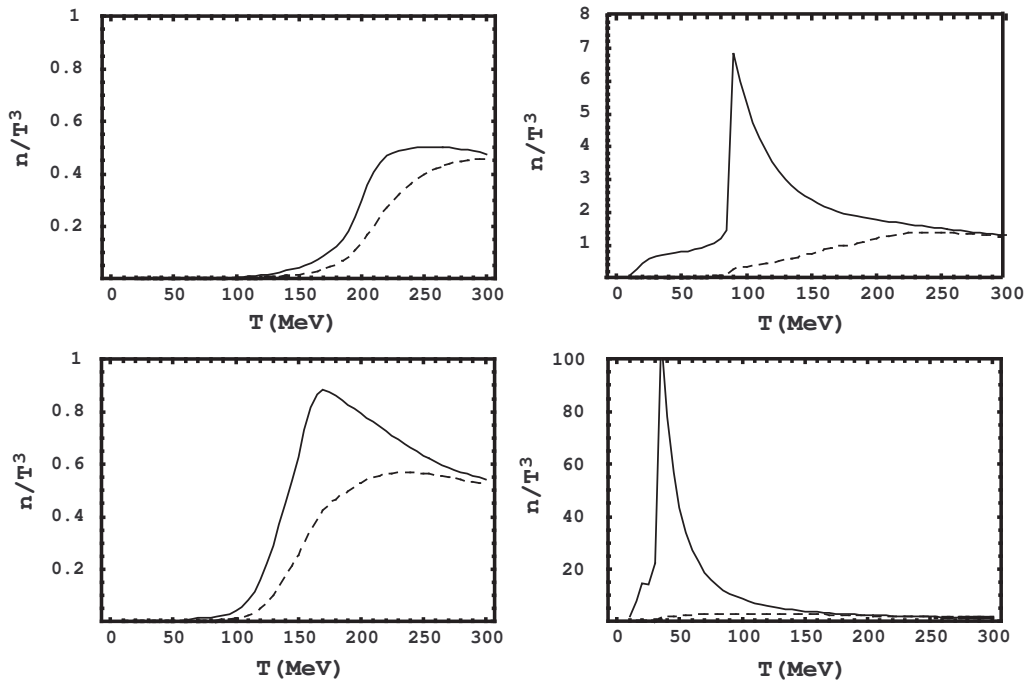}\\
\end{center}
\caption{\label{Fig:numbers} Upper panels: scaled number densities
of quarks $n/T^3$ versus temperature (MeV)  at $\mu=150$ MeV (left)
and $\mu=350$ MeV (right) calculated in the PNJL model with fixed
$T_0 = 270$ MeV; solid lines correspond to up and down quarks,
dashed lines to the strange quark. Lower panels: the same quantities
for the model with $T_0 = T_0(\mu)$.}
\end{figure}

Another possible signature of the deconfinement transition is
offered by the behavior of the quark number densities. In
Fig.~\ref{Fig:numbers} we show the results for the scaled quark
number densities $\tilde{n}_f$. For each flavor $f$, $\tilde{n}_f$ is
defined as follows:
\begin{equation}
\tilde{n}_f \equiv \frac{n_f}{T^3} = -\frac{1}{T^3}\frac{\partial
\Omega}{\partial\mu_f}~.
\end{equation}We plot $\tilde{n}_f$ as a function of $T$
 at $\mu=150$ MeV (left) and $\mu=350$ MeV
(right) for the two cases $T_0=270$ MeV and $T_0(\mu)$. At fixed
quark chemical potential the quark number density is almost
vanishing below the chiral transition temperature, rising quickly
in proximity of the transition itself. This behavior has been
interpreted in the two flavor case as a simulation of the
confined$\rightarrow$deconfined transition in the PNJL
model~\cite{Ratti:2005jh}.

\section{Comparison with nuclear matter}
We shall try here to confront the previous calculations with a
concrete phenomenological model for the transition from nuclear matter
to quark matter. The model will be based on a rather rough
modelisation, but we shall see that the emerging pattern will still
not be very far from that suggested in our previous analysis.
In order to locate more accurately the deconfinement transition,
as well as to study its nature, we now examine the pressure of a
gas made by hadrons at small temperature and small baryonic
chemical potential $\mu_B=3\mu$ and compare it to the pressure of
the PNJL model, the transition occurring when the difference
between the two pressures vanishes. Nuclear matter will be
described by an effective model based on the non-linear extension
of the original Walecka model~\cite{Walecka:1974qa} due to Boguta
and Bodmer~\cite{Boguta:1977xi,Glendenning:1997wn} (WBB in the
following). In its two-flavor version this model allows to predict both the
compression modulus and the effective nucleon mass at saturation
in agreement with their empirical values.
The spinor content of the model we consider here is the whole lightest baryon octet.
We will work assuming $SU(3)$ flavor symmetry for the
couplings, but including mass differences in the baryon octet.

The effective Lagrangian of nuclear matter in the WBB model
describes the baryons coupled to the $\sigma$ scalar meson and the
$\omega$ vector meson, and takes into account cubic and quartic
self-interaction terms for the scalar field:\begin{eqnarray} {\cal
L}&=&\sum_{B}\bar\psi_B\left[i\gamma^\mu(\partial_\mu+i g_{\omega}
\omega_\mu) -(M_B-g_{\sigma}\sigma)\right]\psi_B +{1\over
2}\left(\partial^\mu\sigma\partial_\mu\sigma-
m_\sigma^2\sigma^2\right)\nonumber\\&+& {1\over 2}m_\omega^2
\omega^\mu \omega_\mu- {1\over 4}\omega_{\mu\nu }\omega^{\mu\nu }
-{1\over 3}b_M(g_\sigma\sigma)^3+{1\over 4}c(g_\sigma\sigma)^4~.
\label{eq:nucl2}
\end{eqnarray}
In this equation $\omega_{\mu\nu
}=\partial_\mu\omega_\nu-\partial_\nu\omega_\mu$ is the field
strength tensor of $\omega$. The fields for the $\sigma$ and
$\omega$-mesons are denoted by $\sigma$,
 $\omega_{\mu}$  respectively and $\psi_B$ is the Dirac spinor
for the baryon $B$ having mass $M_B$ (the sum over the index $B$ is
over the entire octet). Moreover $g_{\sigma}$, $g_{\omega}$ are the
dimensionless coupling constants of the $\sigma$ and $\omega$
fields; $b_M$ and $c$ are  parameters; $m_{\sigma}$, $m_{\omega}$
are the masses of the $\sigma$, $\omega$-mesons respectively. Values
of the other parameters are in Table~\ref{tab1}. We use two
parameter sets. Set 1  corresponds to the softest nuclear equation
of state discussed in~\cite{Glendenning:1997wn}; Set 2 to the
stiffest one. These parameters are identical to those used in the
model with nucleons only (without hyperons). As discussed
in~\cite{Glendenning:1997wn} the use of the same parameters is
justified since they are obtained by the fit of some nuclear
properties at the saturation density and zero temperature; in these
conditions the hyperons do not play any role and cannot influence
the result.

\begin{table}
\begin{tabular}{|c|c|c|c|c|}
\hline
 & $b_M/M_N$ & $c$ & $(g_\sigma/m_\sigma)^2$ (fm$^2$) & $(g_\omega/m_\omega)^2$  (fm$^2$)  \\
\hline
Set 1 & $1.46\times 10^{-2}$ & $-1.24\times10^{-2}$ & $9.9262$ & $4.233$ \\
\hline
Set 2  & $2.95\times10^{-3}$ & $-1.07\times10^{-3}$ & $11.79$ & $7.15$ \\
\hline
\end{tabular}
\caption{\label{tab1}Two sets of parameters for the nuclear matter
effective lagrangian, see Eq.~\eqref{eq:nucl2}. We present results
for the dimensionless constant $b_M/M_N$, related to the cubic
interaction coupling constant $b_M$ by the nucleon mass $M_N$.}
\end{table}

We limit ourselves to the mean field approximation,  in which the
meson field operators are replaced by their expectation values;
moreover, the ground state of the system we are interested in
consists of static and uniform matter. This fact implies that the
expectation values of the meson fields do not depend on time and
space coordinates. At finite temperature the equations of motion
for meson and nucleon fields are~\cite{Glendenning:1997wn}
\begin{eqnarray}
&&g_\sigma\sigma
=\left(\frac{g_\sigma}{m_\sigma}\right)^2\left[\frac{1}{\pi^2}
\sum_B \int k^2
\frac{M_B-g_\sigma\sigma}{\sqrt{k^2+(M_B-g_\sigma\sigma)^2}}~f(\epsilon_B(\mathbf{k}), T)~ dk~-b_M(g_\sigma\sigma)^2-c(g_\sigma\sigma)^3 \right],
\label{eq:gsigma}\\
&&g_\omega\omega_0 = \left(\frac{g_\omega}{m_\omega}\right)^2
\rho ,~~~~~~~~~~~~~~~~~~~~~~~ m^{2}_\omega \omega_k = 0,~~~~~~~~~~~~~~~~~~~~~~~~~~~~~~~~~~~~~~~~~~~~~~~~~~~
\label{eq:omega}\\
&&\left[\gamma_\mu(k^\mu -
g_\omega\omega^\mu)-(M_B-g_\sigma\sigma)\right]\psi_B(k)=0~,
\end{eqnarray}
where  $\rho$ is the total baryon density given by \be \rho\equiv
\langle \psi^\dag \psi\rangle =2\sum_B\int
\frac{d\mathbf{k}}{(2\pi)^3}~ f(\epsilon_B(\mathbf{k}),
T)~,\label{eq:BaryonDen}\ee $ f(\epsilon_B(\mathbf{k}), T)=
\displaystyle \frac{1}{1+
\exp\left({\frac{\epsilon_B(\mathbf{k})-\mu}{T}}\right)}$ is the
Fermi distribution function, and $\epsilon_B(\mathbf{k})$ is the
fermion dispersion relation given by\be
\epsilon_B(\mathbf{k})=g_\omega\omega_0 +
\sqrt{k^2+(M_B-g_\sigma\sigma)^2}~.\ee The pressure of  nuclear
matter is given by \bea &p&=-{1\over
3}b_M(g_\sigma\sigma)^3-{1\over 4}c(g_\sigma\sigma)^4 - {1\over 2}
m_\sigma^2\sigma^2 + {1\over 2}
m_\omega^2\omega_0^2\nonumber\\&&~~~~~+\frac{1}{3\pi^2}\sum_B\int
\frac{k^2}{\sqrt{k^2+(M_B-g_\sigma\sigma)^2}}~
f(\epsilon_B(\mathbf{k}), T)~  k^2 dk~. \label{eq:PmuNUCLp}\eea
For the evaluation of the nuclear matter pressure at a given
chemical potential and a given temperature we solve
self-consistently Eqs.~\eqref{eq:gsigma} and~\eqref{eq:BaryonDen},
obtaining the values of $g_\sigma \sigma$ and $\rho$ that are used
in Eq.~\eqref{eq:PmuNUCLp}.

When we compare the pressures of nuclear matter and quark matter,
we face a normalization problem. As a matter of fact, at $T=\mu=0$
the self-consistent solutions of Eqs.~\eqref{eq:gsigma}
and~\eqref{eq:BaryonDen} are $\rho=0$ and $g_\sigma \sigma=0$,
which imply $\omega_0 = 0$ by virtue of the equation of motion
Eq.~\eqref{eq:omega}, and the vanishing of the phase space
integral in Eq.~\eqref{eq:PmuNUCLp}. Clearly quark pressure must
be lower than the nuclear one at relatively low values of density
and/or temperature since in this region of parameters one is in
the confined regime. The quark pressure, which is obtained by the
thermodynamic potential $\Omega$ in Eq.~\eqref{eq:OmegamupNJL}
changing its sign, does not satisfy such a requirement. This
problem is well known in NJL models~\cite{Wang:2006fh}, and the
usual procedure is to subtract a positive constant $B$, the bag
constant, from the quark pressure. The choice of  $B$ is not
unique, and several different suggestions have been analyzed in
the literature~\cite{Wang:2006fh}. We choose to fix the bag
constant by imposing that the transition temperature from nuclear
to quark matter at zero chemical potential coincides with the
crossover temperature $T_\Phi$  of the PNJL model discussed in the
previous Section, i.e. the temperature where $ d\Phi/dt$ has a
maximum. In Table~\ref{tab2} we show the values of
the bag constants and of $T_\Phi$ obtained for the various models.
For each of the quark matter models considered here, the values of $B$ obtained
in comparison with soft and stiff nuclear matter differ
of some part per thousand,
therefore we show only one value of $B$ for each quark matter model.

\begin{table}
\begin{tabular}{|c|c|c|c|c|}
\hline
 & $B^{1/4}$ (MeV) & $T_{\Phi}$ (MeV) & $\mu_c$ stiff (MeV) & $\mu_c$ soft (MeV)\\
\hline
Polynomial, $T_0=270$ MeV  & $428.8$ & 223 & 543 & 617 \\
\hline
Polynomial, $T_0=T_0(\mu)$  & $426.1$ & 160 &  500 &  588 \\
\hline
Logarithmic  & $425.6$  & 153 & 490  & 566 \\
\hline
\end{tabular}
\caption{\label{tab2} Bag constants, deconfinement temperatures at $\mu=0$ and deconfinement chemical potentials at
$T=0$ for the different quark matter
models. Values of $B$ obtained in comparison with soft and stiff nuclear matter differ
of some part per thousand, therefore we show only one value of $B$ for each quark matter model.}
\end{table}

\begin{figure}[t]
\begin{center}
\includegraphics[width=7.7cm]{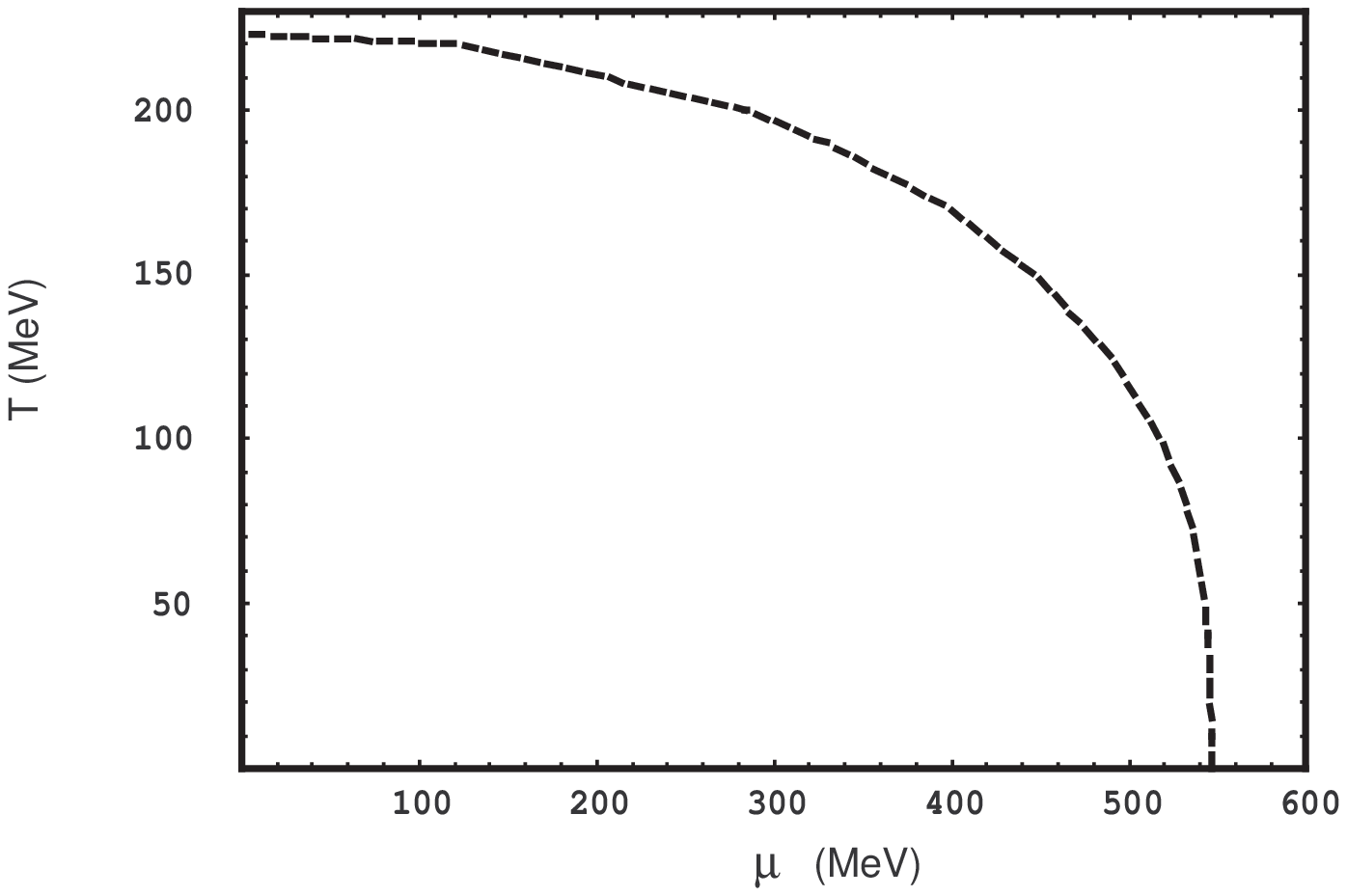}~~
\includegraphics[width=8cm]{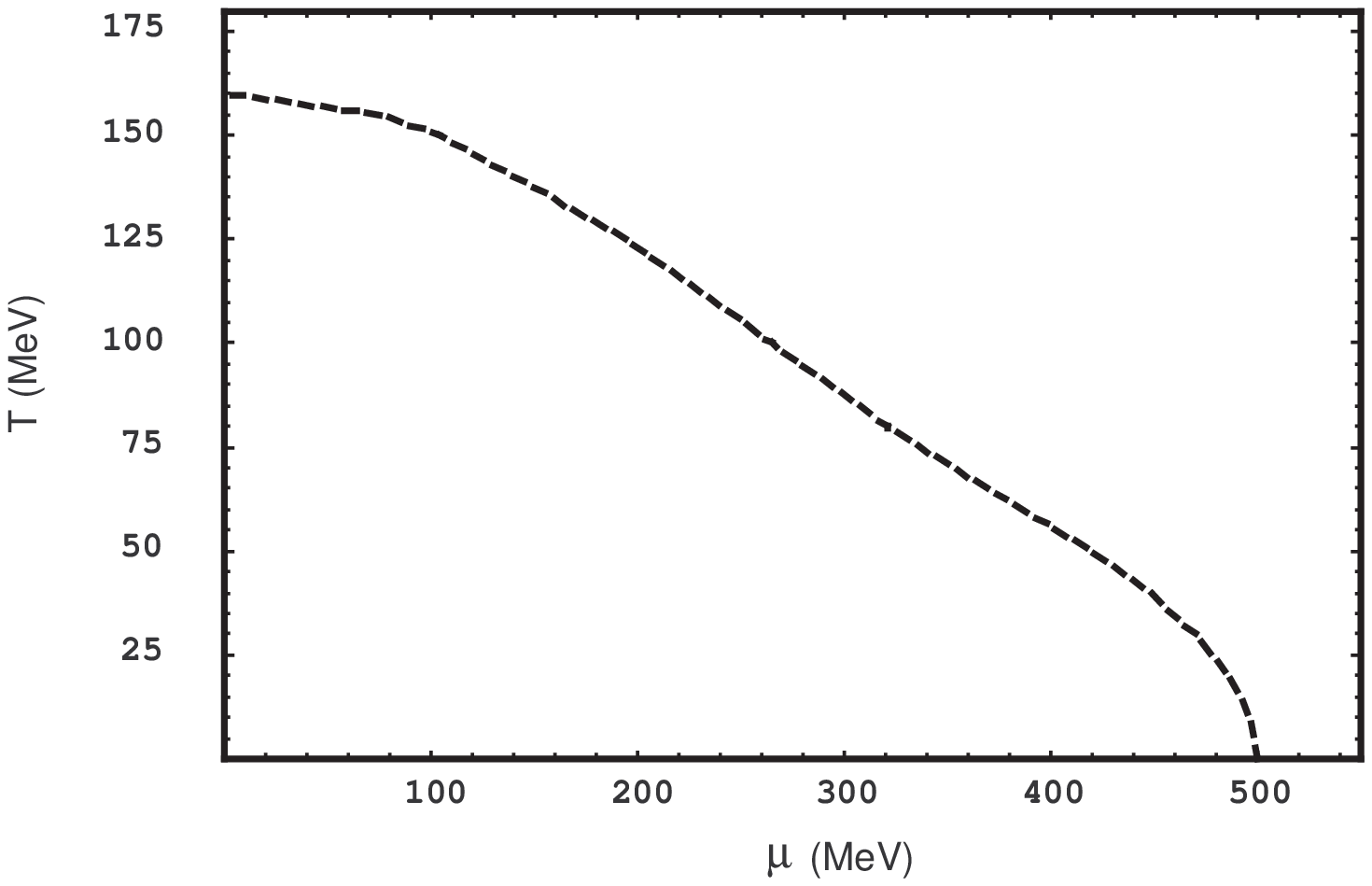}
\caption{\label{Fig:comparison} Nuclear matter/quark matter phase
diagrams in the PNJL model with $T_0 =270$ MeV (left) and $T_0(\mu)$
(right). Nuclear matter contains hyperons and is described by the
stiff equation of motion. Nuclear matter is favored in comparison
with quark matter in the region below the line.}\end{center}
\end{figure}

The results for the two cases are reported in
Fig.~\ref{Fig:comparison}. It shows the phase diagram obtained by
comparing  nuclear (with hyperons) and quark matter. The latter
corresponds to the PNJL model with $T_0 = 270$ MeV (on the left
panel) and $T_0 = T_0(\mu)$ (right panel).  We stress that a more realistic treatment would require the
introduction of different coupling constants in the baryon octet,
but this is beyond the scope of the present study. The nuclear
matter is chosen to be the one with the stiff equation of state (Set
2 of Table~\ref{tab1}). The transitions are of the first order. At
$T=0$ we find the critical quark chemical potential $\mu_c \approx
543$ and $\approx 500$ MeV for the two cases. For the case of the logarithmic potential
we find $\mu_c \approx
490$ MeV.
If we replace the stiff nuclear matter with the soft one, the phase
diagram is qualitatively the same as before; in this case we find
$\mu_c \approx 617$ MeV,  $\mu_c \approx 588$ MeV,  $\mu_c \approx 566$ MeV,  respectively
for the polynomial potential with $T_0 = 270$ MeV, the polynomial potential with
$T_0 = T_0(\mu)$ and the logarithmic potential.

A comparison with Fig. \ref{Fig:diagramma} shows that at small
chemical potential the deconfinement temperature almost coincide
with the chiral restoration temperature for the model with fixed
$T_0$ (left panel in Fig.\ref{Fig:comparison}), while for the
$\mu$-dependent $T_0$ (right panel in Fig.\ref{Fig:comparison}) the
two temperatures differ. Moreover, at small temperatures the
critical baryonic chemical potential is higher for the deconfinement
transition. However the values of the chemical potential found here
are such that one of the color superconductive phases might be
energetically favored~\cite{CSC,Ruster:2005jc,Blaschke:2005uj,Ippolito:2007uz},
a possibility we have not considered in the
present paper for simplicity.

\section{Conclusions}
We have studied the phase diagram of the PNJL model with three
flavors, considering different versions of the model, i.e. with
polynomial and logarithmic forms of the Polyakov loop potential, in
the former case with two forms of the reference temperature: $T_0$
fixed (=270 MeV) and $\mu$-dependent $T_0(\mu)$. One result is
related to the chiral symmetry phase transitions and is plotted in
Fig.~\ref{Fig:diagramma}. In general we find at $\mu=0$ crossover
temperatures higher than the analogous result of the NJL model,
namely $T=175$ MeV. Moreover we find that the critical endpoint
slightly depends on the choice of the model, but for all the cases
considered in this paper the value of $\mu_E$ is higher than the
result obtained by lattice QCD or QCD-like theories ($\mu_E\,\sim$
50 MeV). The PNJL result might be improved by adding a new
interaction term ${\cal O}(\bar\psi \psi)^4$ in the quark lagrangian
as in Ref.~\cite{Kashiwa:2007hw}. We leave this investigation to a
future project.

We have also studied the role of the Polyakov loop in the
nuclei$\rightarrow$quarks transition. In order to describe nuclear
matter we have adopted the improved Walecka model with a
self-interacting scalar field $\sigma$ and a vector field $\omega_\mu$. We
have chosen two sets of parameters of the nuclear matter lagrangian,
corresponding respectively to a soft and to a stiff equation of
state \cite{Glendenning:1997wn}. We find similar results in the two
cases. The result of this analysis is summarized in
Fig.~\ref{Fig:comparison} (stiff equation of state). The quark
pressure is normalized by imposing that the deconfinement crossover of the PNJL model
and the nuclei$\rightarrow$quarks transitions occur at the same temperature
at $\mu=0$.  We get
that chiral symmetry restoration temperature and deconfinement
temperature almost coincide at $T=0$ in the case of fixed $T_0$ and
polynomial form of the Polyakov loop  potential, while for the
$\mu$-dependent $T_0$ the two temperatures differ. We also find that
at small temperatures the critical chemical potential for deconfinement is higher
than the one for the chiral transition, see Figs.~\ref{Fig:diagramma} and~\ref{Fig:comparison}.

This work might be improved in several different ways. For example
one should impose  electric and color neutrality in the quark matter
sector. Also the possibility of a color superconductive phase at
high density should be taken into account. Moreover the nuclear
matter equation of state should be replaced by a more sophisticated
one taking into account population imbalances due to
$\beta-$equilibrium and electric neutrality. We leave all these
issues to future projects. We finally note that while preparing this text
we became aware of an almost simultaneous and independent study of the three
flavor PNJL model~\cite{new} that has some overlap with our work.

\acknowledgements

We thank L.~Cosmai, K.~Fukushima, G.~Pulvirenti and B.~J.~Schaefer for useful discussions.

\end{document}